# Long-term variations of quasi-trapped and trapped electrons in the inner radiation belt observed by DEMETER and SAMPEX

Kun Zhang[1,2], Xinlin Li[1,2], Zheng Xiang[2,3], Leng Ying Khoo[1,2], Hong Zhao[2], Mark D. Looper[4], Michael A. Temerin[5], and Jean-André Sauvaud[6]

[1]Department of Aerospace Engineering Sciences, University of Colorado Boulder, Boulder, Colorado, USA.

[2]Laboratory for Atmospheric and Space Physics, University of Colorado Boulder, Boulder, Colorado, USA.

[3]Department of Space Physics, School of Electronic Information, Wuhan University, Wuhan, Hubei, China.

[4]Space Sciences Department, The Aerospace Corporation, El Segundo, CA, USA.

[5](retired from) Space Sciences Laboratory, University of California, Berkeley, California, USA.

[6]Institut de Recherche en Astrophysique et Planétologie-IRAP, Université de Toulouse, CNRS, UPS, CNES, Toulouse, France

Kun Zhang (Kun.Zhang@lasp.colorado.edu)

**Key Points:**

1. Sub-MeV electrons at L≤1.14 are anti-correlated with sunspot number suggesting their source to be cosmic ray albedo neutron decay.
2. Electrons at L≥1.2 are enhanced during large geomagnetic storms and decay to a background level during extended quiet times.
3. Quasi-trapped electrons at L>2 are highly correlated with trapped electrons, indicating that pitch angle scattering dominates.

**Abstract**

Electrons in the Earth's radiation belts can be categorized into three populations: precipitating, quasi-trapped and trapped. We use data from the DEMETER and SAMPEX missions and from ground-based neutron monitors (NM) and sunspot observations to investigate the long-term variation of quasi-trapped and trapped sub-MeV electrons on different L shells in the inner belt. DEMETER and SAMPEX measurements span over 17 years and show that at L≤1.14 the electron flux is anti-correlated with sunspot number, but proportional to the cosmic ray intensity represented by NM count rates, which suggests that electrons at the inner edge of the inner belt are produced by Cosmic Ray Albedo Neutron Decay (CRAND). The solar cycle variation of cosmic rays increased the electron flux at L≤1.14 by a factor of two from solar maximum at 2001 to solar minimum at 2009. At L≥1.2, both quasi-trapped and trapped electrons are enhanced during geomagnetic storms and decay to a background level during extended quiet times. At L>2, quasi-trapped electrons resemble trapped electrons, with correlation coefficients as high as 0.97, indicating that pitch angle scattering is the dominant process in this region.

**1 Introduction**

1.1 Radiation Belt Electrons

Earth's radiation belts are subject to both transient phenomena and long-term variations. The outer electron belt is most dynamic, often enhanced or decreased by geomagnetic storms whose effect can last for several days. In contrast, the inner electron and proton belts are more stable. In general, energetic protons in the radiation belt, which are known to be mostly created by Cosmic Ray Albedo Neutron Decay (CRAND), are anti-correlated with sunspot number, though occasionally increased by solar energetic proton events (Baker et al., 2004; Li et al., 2001; Miyoshi et al., 2000; Selesnick et al., 2010). On the other hand, the electron flux in the inner belt typically decays steadily, except during extremely large geomagnetic storms when electrons are transported into the inner belt (e.g., Li et al., 2015; Selesnick, 2016a; Zhao & Li, 2013). Sources of inner belt electrons include inward diffusion (e.g., Cunningham et al., 2018; Selesnick, 2016a) and enhanced large-scale electric fields (Selesnick, 2016b; Su et al., 2016). In addition, Li et al. (2017) identified CRAND as an important source of inner belt electrons in certain regions, especially at the inner edge of the inner belt.

A major loss process for radiation belt electrons is precipitation into the Earth's atmosphere, which is the dominant loss process at low L (L can be viewed as the geocentric distance in $R_E$ at the geomagnetic equator if the geomagnetic field is approximated as a dipole). However, at L>1.6, radiation belt electrons generally precipitate through pitch angle scattering by various waves in the magnetosphere, including whistler-mode waves and electromagnetic ion cyclotron (EMIC) waves (Blum et al., 2015; Li and Hudson, 2019). Enhanced electron loss due to wave-particle interaction may efficiently slow the increase of electron intensity and even decrease it (e.g., Xiang et al., 2018). On the other hand, at the inner part of the inner belt (L=1.2 to 1.6), atmospheric collisions become the main mechanism inducing pitch angle scattering (Selesnick, 2012; Xiang et al., 2020).

Because the magnetic field is weak near the South Atlantic Anomaly (SAA), the bounce loss cone increases near the SAA and some electrons that survive in other regions precipitate when they drift into the SAA. Such electrons are in the drift loss cone and are usually referred to as "quasi-trapped" electrons. Quasi-trapped electrons are observed in the inner belt, slot regions and outer belt (e.g., Tu et al., 2010; Zhang et al., 2017). Electrons that are trapped at all longitudes including the SAA region are considered "stably-trapped" or just "trapped". Trapped electrons can accumulate through multiple drift periods and thus usually have higher fluxes compared to the quasi-trapped electrons. However, they may be scattered into the bounce loss cone, becoming "precipitating" electrons that are immediately lost to the atmosphere or they may be scattered into the drift loss cone, becoming quasi-trapped electrons. Li et al. (2017) showed that trapped electrons only exist at L>1.14 and quasi-trapped electrons observed at L<1.14 are produced by CRAND. More detail about how to distinguish these populations from each other is in Section 2.3.

1.2 Cosmic Rays

Cosmic rays consist of high energy protons and heavier nuclei, probably energized by supernovas within our galaxy (Blandford & Eichler, 1987). Because of shielding by the magnetic fields of the Sun and of the Earth, some cosmic rays are not energetic enough to reach near-Earth space even if they travel in the direction of Earth. The cosmic rays that reach the Earth's upper atmosphere interact with atmospheric atoms and can generate neutrons through two mechanisms (Ifedili et al., 1991). If the kinetic energy of the incident cosmic ray is above ~10 MeV, the cosmic ray can strike a neutral atom and release neutrons that carry most of the cosmic ray's momentum, which is called the knock-on process. Therefore, knock-on neutrons may have energies greater than 10 MeV and travel in almost the same direction as the incident cosmic ray. On the other hand, cosmic rays may also excite atomic nuclei and cause them to release neutrons, which is called the evaporation process. Evaporative neutrons have relatively low energies, peaked at 1 MeV, and make up about 90% of all cosmic ray produced neutrons (Hess et al., 1961; Simpson, 1951).

Neutrons have an average lifetime of 887 s and decay into protons, electrons and antineutrinos. About 10% of cosmic ray neutrons travel upward from the atmosphere and decay into protons and electrons that can be trapped in the Earth's magnetosphere (Ifedili et al., 1991), a process known as Cosmic Ray Albedo Neutron Decay (CRAND). Figure 1a is an illustration of different kinds of CRAND processes, showing that albedo neutrons can travel a certain distance before they decay. High energy neutrons mostly originating from the knock-on process travel fast approximately in a straight line and can decay either in or out of the Earth's magnetosphere. In contrast, some low energy neutrons from the evaporation process fail to escape Earth's gravity and can decay at relatively low altitudes. The decay of knock-on neutrons is believed to be the source of the radiation belt protons (Singer, 1958, 1962), whereas the evaporation process has recently been identified as the source of electrons at the inner edge of the inner belt (Li et al., 2017; Xiang et al., 2019; Zhang et al., 2019).

Cosmic rays reaching near-Earth space can be measured indirectly by ground-based neutron monitors and by particle detectors on low altitude spacecraft as illustrated in Figure 1a. Ground-based neutron monitors are distributed across the globe, and early studies showed that cosmic ray intensity reaches minimum near the equator and maximum near the geomagnetic poles (Compton and Turner, 1937; Rose et al., 1956). Because of shielding by Earth's magnetic field, cosmic rays have more access to open field lines near the poles and less access near the geomagnetic equator. This feature is usually discussed using the cut-off rigidity, which is the lowest rigidity or energy for a cosmic ray to arrive at a specific location on Earth. A series of studies have calculated and characterized the cut-off rigidity (Shea et al., 1965; Shea, 1969) and it was concluded that the vertical cut-off rigidity (R), the cut-off rigidity for cosmic rays arriving from the vertical direction, is approximately $R(GV)=14.9/L^2$ (Shea, 1987) where L is the McIlwain L (McIlwain, 1961). Figure 1b is calculated based on this relationship, where the rigidity is shown as contours with the corresponding L shell noted in the color bar. Figure 1b shows that cosmic ray protons need to have energies of at least several GeV in order to reach the inner belt region. However, since neutrons can travel some distance ignoring the magnetic field before decay, cosmic rays with slightly lower energies could also contribute to the inner belt. Recent missions such as PAMELA and AMS provide direct measurements of cosmic rays at the top of the atmosphere (Adriani, 2011; Aguilar, 2015). Both ground-based neutron monitors and PAMELA observed that cosmic ray intensities have a solar cycle dependence which is anti-correlated with the sunspot number and delayed in phase by less than one year compared to the sunspot number (Adriani, 2013; Inceoglu, 2014; Martucci, 2018).

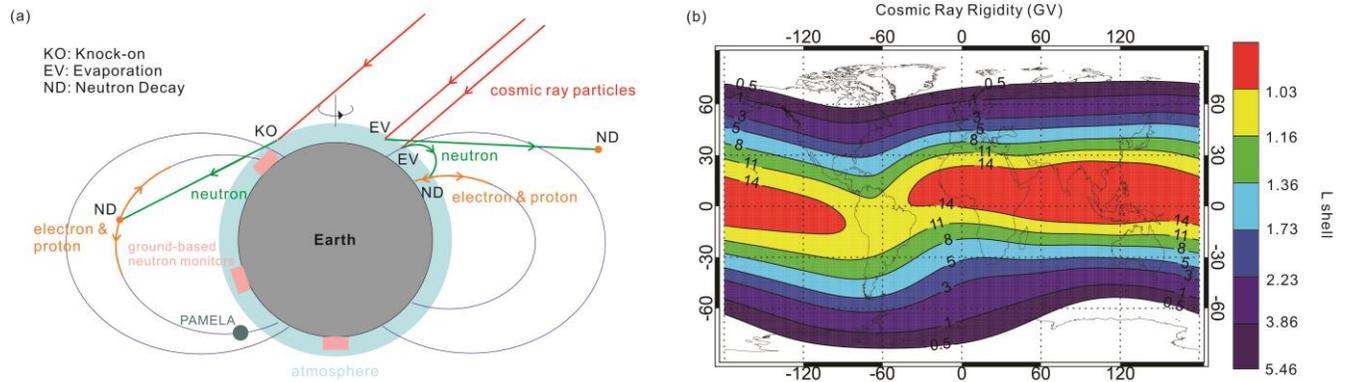

**Figure 1**. (a) Illustration of different mechanisms of CRAND including the knock-on process, the evaporation process and the decay of neutrons. Blue curves represent geomagnetic field lines. Trajectories of different particle species are shown in different colors and examples of locations where cosmic ray measurements take place are shown as the ground-based neutron monitors and PAMELA (satellite). (b) Calculated cosmic ray vertical cut-off rigidity in GV shown as colored contours on a map, with the color bar indicating the corresponding L shell.

In this paper, we use measurements from DEMETER and SAMPEX to investigate the long-term variations of quasi-trapped and trapped sub-MeV electrons in the inner radiation belt. We also include some cosmic ray observations for comparison. We show the different features of electrons at different L, which imply different source and loss mechanisms. Finally, we quantify the observed relationships via statistical analysis.

**2 Data and Methodology**

2.1 Electron Measurements

DEMETER (short for "Detection of Electro-Magnetic Emissions Transmitted from Earthquake Regions") had a sun-synchronous circular orbit of 710 km altitude and 98.3° inclination. The Instrument for the Detection of Particle (IDP) onboard provided electron flux measurement from 70 keV to 2.4 MeV at a fine energy resolution (Sauvaud et al., 2006). The routine mode data have an ~18 keV energy resolution and a 4 s cadence. Sauvaud et al. (2006) showed that DEMETER/IDP had more accurate measurements in the sub-MeV range, which is the population on which we focus. Data from 2004 to 2010 are used in this study.

The Solar, Anomalous, and Magnetospheric Particle Explorer satellite (SAMPEX) was also a polar-orbiting low-altitude satellite, carrying the Proton/Electron Telescope (PET) (Cook et al., 1993). PET measured electron and proton intensities from 1992 to 2009. Selesnick (2015) revisited the response function of PET and determined that the P1 channel (lowest energy channel) of PET measures >360 keV electrons. The instrument's live time is used to describe the interval during which the incident particles can be recorded and is decreased during high flux levels resulting in the underestimate of fluxes (Selesnick, 2015). Therefore, here we apply a live time correction to the electron flux measurements and the corrected flux $J_c$ is

$$J_c = J \times \frac{586}{livetime - 150}$$

where J is the original flux. The look directions of both the SAMPEX and DEMETER usually pointed roughly perpendicular to the local magnetic field in the radiation belts, observing locally mirroring particles. However, SAMPEX spun at 1 RPM during several periods, such as 3/5/1996 - 3/8/1996, during which time PET also measured particles with small local pitch angles resulting in a lower average flux. Therefore, in this study, we adjusted the fluxes measured in the spinning mode up to the flux level in the normal periods by applying a multiplicative factor to the data. The factors used will be noted where the data are shown. In addition, SAMPEX was in an elliptical orbit around 690×510 km in altitude at the start of the mission, but the orbit had decayed to around 490×410 km by 2009. In order to eliminate the altitude influence on the flux measurements, for each orbit we find the point with altitude closest to 500 km for each day and only include the points within a ±20 km altitude window around that point in the following study.

2.2 Cosmic Ray Measurements

Ground-based neutron monitors have been built across the world and many have been in operation for decades, making them ideal for long-term studies. Here, we select the daily neutron count rate measured at the Mexico City Cosmic Ray Observatory, located at (19.33° N, 260.82° E) and 2.3 km in altitude. This neutron monitor is at L~1.3, corresponding to the inner belt region, and has a vertical cut-off rigidity of 8.2 GV. It has been operating since 1990. In addition, we also use the cosmic ray intensities measured in space by PAMELA, a particle detector onboard the Russian Resurs-DK1 satellite launched in 2006. The satellite was launched into a 350 km × 600 km (altitude) orbit with 70° inclination and later changed to a circular orbit at 580 km altitude in 2010. PAMELA provided >80 MeV proton measurements from 2006 to 2016, with the kinetic energy calculated based on the cut-off rigidity (Adriani et al, 2013; Martucci et al., 2018).

2.3 The Determination of Electron Populations in the Radiation Belts

LEO satellite measurements are significantly affected by local geomagnetic anomalies, and different electron populations in terms of the trapping status are measured at different geographic locations. For example, such satellites observe trapped electrons with high fluxes near the SAA and low precipitating fluxes in the northern hemisphere conjugate to the SAA. Here we adopt the method used in Selesnick (2015) to automatically identify electron populations based on the location of the measurement. We assume that the electrons are lost into the atmosphere when they reach 100 km altitude. First, we find the magnetic field at the electron's mirror point, $B_m$ (for locally mirroring particles, it is simply the local geomagnetic field) and the magnetic fields at the two points with 100 km altitude along the field line that the particle is on, $B_{n100}$ in the northern hemisphere and $B_{s100}$ in the southern hemisphere. Then for each L, we use all the data in that year and find $B_{100}$, defined as the lowest value among all $B_{n100}$ and $B_{s100}$, which describes the drift loss cone. We determine the electron population that is measured at a certain data point using the following conditions: (1) $B_m > B_{n100}$ or $B_m > B_{s100}$: precipitating; (2) $B_m < B_{n100}$ and $B_m < B_{s100}$ and $B_m > B_{100}$: quasi-trapped; (3) $B_m < B_{100}$: trapped. Figure 2 shows an example of the calculation result for DEMETER. The magnetic fields used in the calculation for both satellites are based on the International Geomagnetic Reference Field (IGRF) model (Langel, 1992; Thébault et al., 2015). Simply assuming that a satellite observes locally mirroring particles, instead of using the instrument's actual look direction to determine the pitch angle, will only have minor influence on the results, since both satellites point mostly perpendicularly to the local magnetic field and the sensors have large fields of view. In this study, we use the actual look direction of DEMETER for calculations and assume measuring mirroring electrons for the SAMPEX calculation.

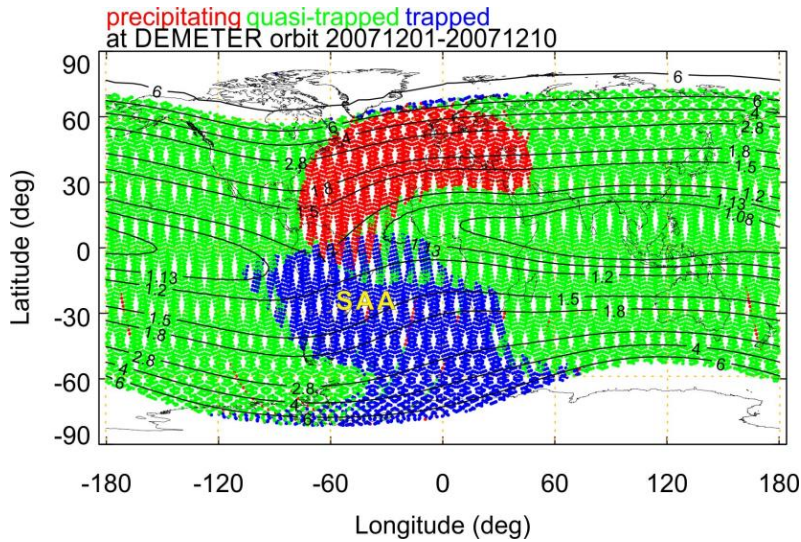

**Figure 2.** DEMETER orbit from Dec 1, 2007 to Dec 10, 2007 plotted over a map color-coded with the populations of the electrons measured by the satellite. L contours are shown as black solid curves. The location of the SAA is marked by a yellow label.

**3 Observations and Discussion**

3.1 Electrons at the inner edge of the inner belt

In this section, we discuss the long-term variation of electrons at the inner edge of the inner belt (L<1.14) both qualitatively and quantitively using measurements from DEMETER and SAMPEX. Figure 2 shows that at L=1.13 the vast majority of LEO satellite measurements are quasi-trapped electrons. We apply the method described in Section 2.3 to select all the quasi-trapped electron measurements at L=1.13-1.14 from DEMETER (2004-2010) and SAMPEX (1993-2009) and calculate the average flux in a running 100-day window, as shown in Figure 3. SAMPEX measurements (blue) are from the P1 channel with energies of >360 keV and selected in altitude close to 500 km, corresponding to the left axis. Due to the altitude filter we applied to SAMPEX data, about 60% of the days in the first 5 years and about 1/3 of the days in the later years have no qualified quasi-trapped electron measurements. SAMPEX data in the spinning periods (shown as shaded area) are multiplied by 1.27 for periods before 2001 and 1.55 for periods after 2001, in order to match the count rate level in the routine periods. DEMETER measurements (red) are from the 500 keV channel and correspond to the right axis in red. The neutron count rate measured in Mexico City, Mexico and the sunspot number are also averaged in a running 100-day window and shown for comparison.

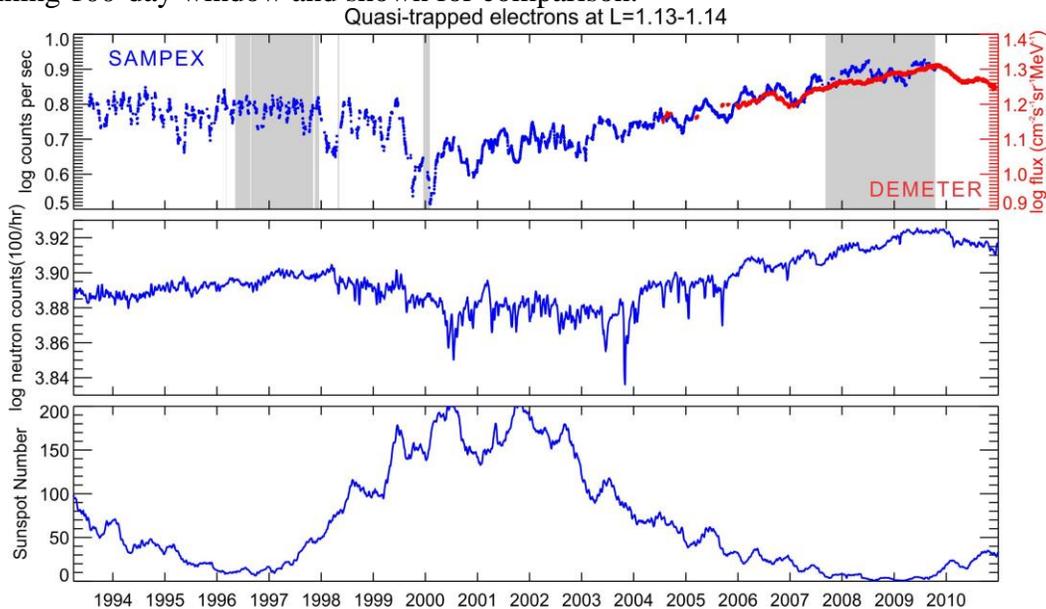

**Figure 3.** Long-term profile of quasi-trapped electrons at the inner edge of the inner belt compared with cosmic ray and solar cycle information from 1993 to 2010. (top) Quasi-trapped electron intensities at L=1.13-1.14 measured by SAMPEX (>360 keV, blue and left axis) and DEMETER (500 keV, red and right axis). Shaded areas indicate periods when SAMPEX was spinning at 1 RPM, and the electron count rate is adjusted with factors of 1.27 for the periods before 2001 and 1.55 for periods after 2001. (middle) Neutron count rate measured at Mexico City, Mexico, corresponding to a cosmic ray cut-off rigidity of 8.2 GV, equivalent to 7.3 GeV/n in kinetic energy. (bottom) Sunspot number representing solar cycle information. Data shown in all three panels are averaged in a running 100-day window.

In Figure 3, quasi-trapped electron fluxes from DEMETER and SAMPEX both generally follow the trend of cosmic ray intensity represented by the neutron count rate and are anti-correlated with the sunspot number, showing high electron fluxes at solar minimum and low electron fluxes at solar maximum, consistent with the long-term variation of CRAND. From solar maximum at 2001 to solar minimum at 2009, the electron count rate almost doubled as observed by SAMPEX. In the period when both DEMETER and SAMPEX data are available, the electron profiles from both satellites are similar, qualitatively validating each other's measurements. Note that a quasi-periodic variation with a period of about one year is observed in the electron flux measured by SAMPEX, which is likely due to the orbital effect: higher electron flux was observed when SAMPEX's altitude above the west of SAA was higher, and vice versa. This variation is not evident for DEMETER measurements because DEMETER has a circular orbit.

The sunspot number generally decreases from about 2001 to 2009, marking the declining phase of the solar cycle from the solar maximum to the solar minimum. During this period, the solar magnetic field decreases, therefore, the shielding effect of the solar magnetic field on the incoming cosmic rays is also weakened, allowing more cosmic rays entering the inner heliosphere and thus near-Earth space. The increased cosmic rays arriving at the Earth's upper atmosphere lead to stronger CRAND process, producing more sub-MeV electrons. Furthermore, a portion of the enhanced CRAND-produced electrons are trapped or quasi-trapped in the Earth's magnetosphere and cause the increase in the electron flux measurements at the inner edge of the inner belt, as observed in Figure 3.

From the end of 2006 to the end of 2009, electron fluxes from both satellites increased by about 25%. However, this does not agree with the percentage change of the neutron count rate which is only about 4%. The differences could be due to either the uncertainty in the background level of the neutron count rate or different altitudes of the measurements. To verify this, we also show the proton component in cosmic rays measured by PAMELA at LEO from 2006 to 2014 in Figure 4, which is also anti-correlated with the solar cycle. Protons in the energy range of 6.99-7.74 GeV/n are close to the population measured by the neutron monitor in Mexico City (7.3 GeV/n) but increased by about 15% from 2006 to 2009 as measured by PAMELA, which is closer to the percentage increase of the electron fluxes measured by SAMPEX and DEMETER. Additionally, as shown in Figure 4, cosmic rays with lower energy experience even larger variation. Even though the lower energy cosmic rays cannot directly access such low L, the produced neutrons can still travel a certain distance and decay into electrons at the inner edge of the inner belt. Taking these uncertainties into consideration, the percentage change of the cosmic ray intensities may be comparable to that of the quasi-trapped electron flux. It is also worthy to mention that the highest cosmic ray intensity since the start of the space age was recorded at the solar minimum near 2009 (Adriani et al., 2013), at which time quasi-trapped electrons at L=1.13-1.14 also had the highest flux in Figure 3. We suggest that detailed simulation focusing on the electron production of the interaction between the cosmic rays and atmosphere needs to be conducted to obtain a comprehensive understanding of the CRAND effect on radiation belt electrons.

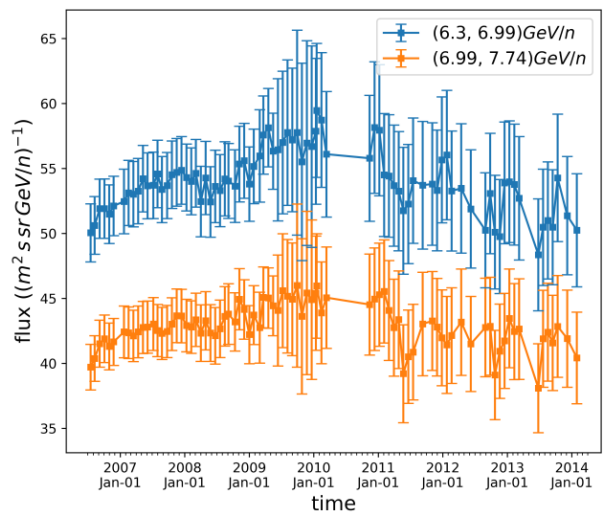

**Figure 4.** Proton fluxes measured by PAMELA from 2006 to 2014 in kinetic energy ranges of 6.3-6.99 GeV/n (blue) and 6.99-7.74 GeV/n (orange). Vertical lines show the error bar of the fluxes. PAMELA data plotted here are directly from https://tools.ssdc.asi.it/CosmicRays.

We further evaluate the above relationships statistically. We take the same data used for Figure 3 and bin them into 14-day windows (for DEMETER) and 50-day windows (for

SAMPEX). SAMPEX data need a larger window because fewer data points are left after we filter them by altitude. Note that for both satellites, if there are less than 1000 data points in a certain time window, this period is considered insufficient for statistics and discarded (10 out of 162 periods are removed for DEMETER, none for SAMPEX). The correlation between the electron fluxes from both satellites and the neutron monitor count rate are shown in Figure 5 (a) and (b). Electron fluxes from both satellites are shown to fit well with the neutron count rates on a linear scale, with a correlation coefficient of 0.71 for DEMETER and 0.64 for SAMPEX, indicating that the electron variation at the inner edge of the inner belt is almost proportional to the variation of cosmic rays. The obvious outlier in Figure 5(a) corresponds to a period around 07/21/2005. In addition to this point, there are 4 points with significantly higher fluxes falling above the y axis range and these 4 points are not included in the calculation of the correlation coefficient. All these 5 outliers are during the extreme geomagnetic activities from Nov 2004 to July 2005, bringing various uncertainties to the measurements. These high flux points also cause the large data gaps shown in DEMETER data in Figure 3, where the 100-day average fluxes are abnormally high and fall beyond y axis range. Furthermore, Figure 3 shows that both electrons and neutrons reach maxima at around the end of 2009 while the sunspot number reaches its minimum at the beginning of 2009, indicating that like neutrons, the long-term electron variation is also out of phase with the actual solar cycle, falling behind by just less than one year. We applied various time lags to the DEMETER electron measurements and calculated the correlation coefficients between the electron flux and the sunspot number. We find that the highest correlation coefficient is 0.67, reached with a time lag of 297 days (shown in Figure 5(c)). In comparison, we use the same method and determine the time lag with highest correlation coefficient using the neutron count rate to be 324 days (shown in Figure 5(d)), which is close to the results from DEMETER. The profiles of the correlation coefficients as a function of time lag are shown in Figure S1 in the supporting information. Inceoglu et al. (2014) calculated the time lag of cosmic rays in this solar cycle to be 8~12 months using other neutron monitors, which is comparable to our results. This demonstrates the close relationship between these electrons and cosmic rays from another perspective. To conclude this section, the high correlation between the quasi-trapped electrons and the cosmic ray intensity suggests that CRAND is the dominant source of electrons at the inner edge of the inner belt.

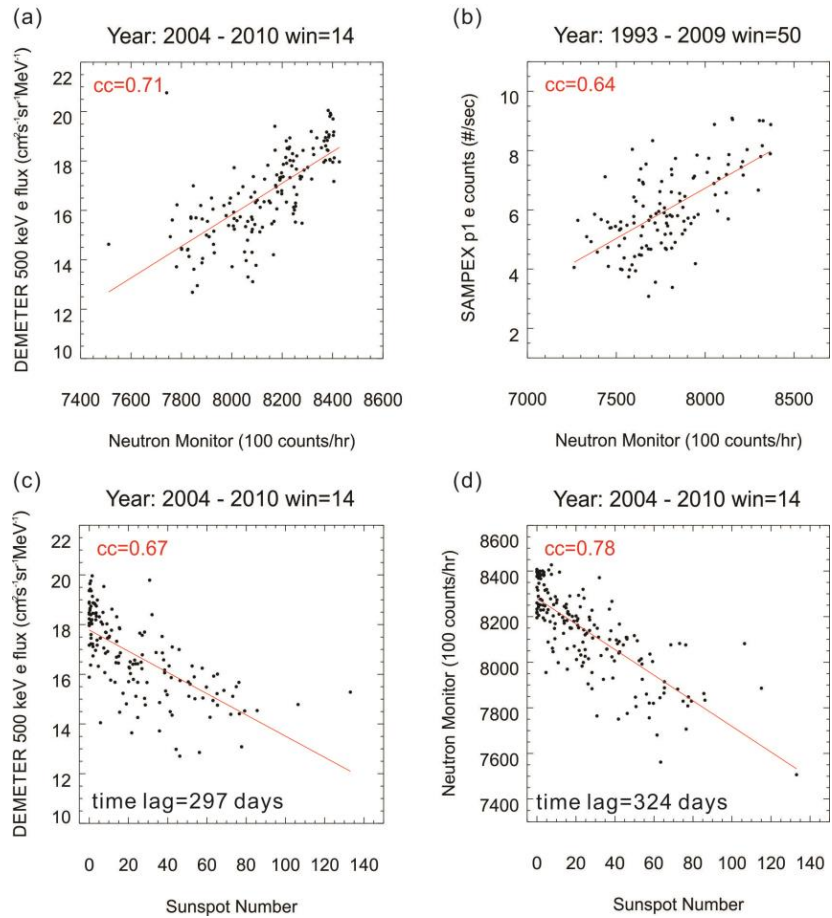

**Figure 5.** Correlations between the quasi-trapped electrons at L=1.13-1.14, the neutron monitor and the sunspot number. (a) DEMETER measured 500 keV electron flux vs. the neutron count rate from 2004-2010. (b) SAMPEX measured >360 keV electron count rate vs the neutron count rate from 1993-2009. (c) DEMETER measured 500 keV electron flux vs. the sunspot number from 2004-2010 with a time lag of 297 days added to DEMETER data. (d) The neutron count rate vs. the sunspot number from 2004-2010 with a time lag of 324 days added to the neutron count rate. Data in (a), (c) and (d) are averaged in 14-day bins and data in (b) are averaged in 50-day bins. The average values are plotted in black dots and fitted to straight lines shown in red. The correlation coefficients are noted in red at the upper-left corner of each panel. (a) and (c) have 152 points in each plot. (b) has 124 points and (d) has 180 points.

3.2 Electrons at higher L shells in the inner belt and slot region

At higher L shells, quasi-trapped and trapped electrons are both abundant and more dynamic, with prompt responses to geomagnetic activity. SAMPEX measurements show that at L=1.20-1.21, both quasi-trapped and trapped electrons are significantly enhanced during large geomagnetic storms especially during the declining phase of the solar cycle when high-speed solar wind is dominating (Figure 6), which is in strong contrast with the behavior of electrons at L≤1.14, where electrons are not affected by geomagnetic storms and are only subject to a gradual solar cycle variation.

It is worth noticing that during the solar maximum from 1999 to 2002, the trapped electron flux is not significantly enhanced even though many storms happened during that time. In contrast, electron flux increased rapidly at the start of 2003. This is consistent with the observations at geosynchronous orbit. For example, Borovsky and Denton (2006) showed that the high fluxes of relativistic electrons were predominantly associated with Corotating Interaction Regions (CIR)-driven storms and occurred mostly during the declining phase of the solar cycle (1993-1996), while only one event with high fluxes during 1996-2002 was related to Coronal Mass Ejections (CME), which generally drives many storms during the solar maximum. Zheng et al. (2006) also studied the same period using SAMPEX measurements and concluded that flux enhancements in the inner belt are associated with high solar wind speed. Similar study by Li et al (2011) also concluded that the averaged flux of relativistic electrons is higher during the declining phase than during the maximum of solar cycle. In addition, trapped electrons at L=1.2 have long lifetime, therefore, when several injections happened even with many days apart, the electron flux can stay elevated.

Quasi-trapped electrons generally follow the trend of trapped electrons except for the difference in the flux magnitude and that the quasi-trapped electron fluxes decrease faster after the storm ends. During extended quiet times such as 2006-2009, quasi-trapped electrons decay to a low background flux level, which is likely maintained by CRAND (Xiang et al., 2020; Zhang et al., 2019). Similar results are also observed by DEMETER (Figure 7 (a) and (b)). Figure 7 (c) and (d) show that quasi-trapped and trapped electrons at L=2 are more dynamic than those at L=1.2, frequently enhancing and decreasing. Frequent enhancements indicate that injected electrons can access L=2 more easily, while the faster declines indicate that pitch angle scattering is more efficient at L=2 due to the existence of various wave-particle interactions (Ripoll et al., 2019).

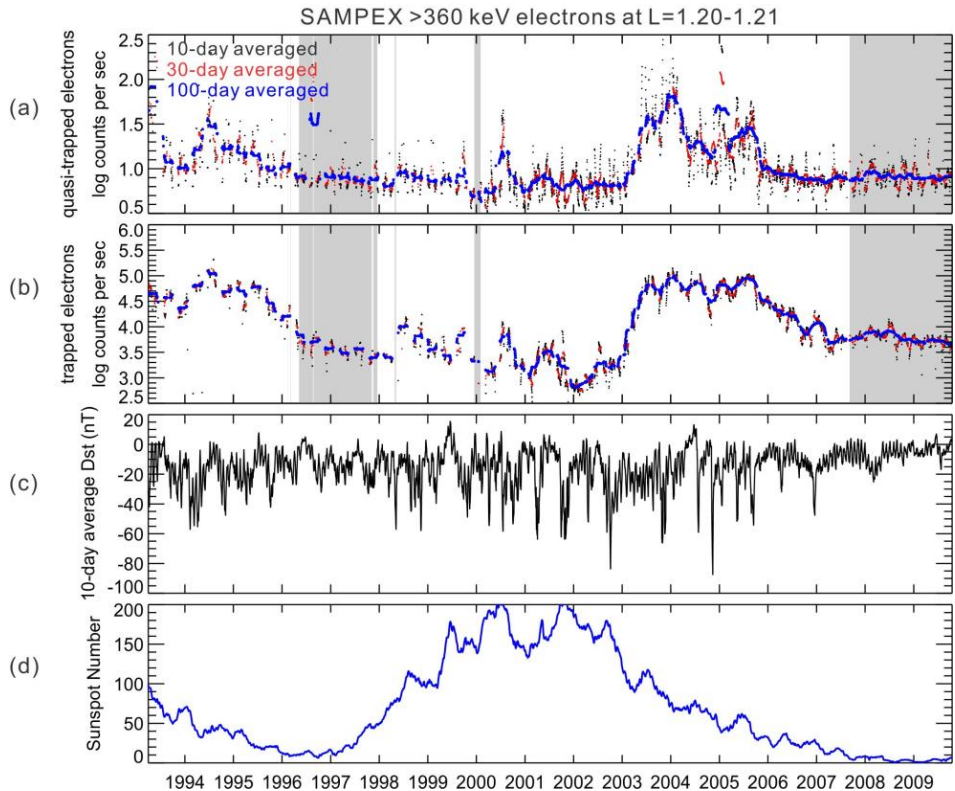

**Figure 6.** Profile of >360 keV quasi-trapped and trapped electrons measured by SAMPEX at L=1.20-1.21 compared with Dst and the Sunspot number. Data are averaged in 10-day (black), 30-day (red) and 100-day (blue) running windows. (a) Quasi-trapped electrons measured by SAMPEX with the count rate in the spinning periods (shaded area) adjusted by factors of 1.26 before 2001 and 1.52 after 2001. (b) Similar to (a) for trapped electrons with the adjustment factors of 0.92 before 2001 and 2.47 after 2001. (c) 10-day averaged Dst index. (d) 100-day averaged sunspot numbers.

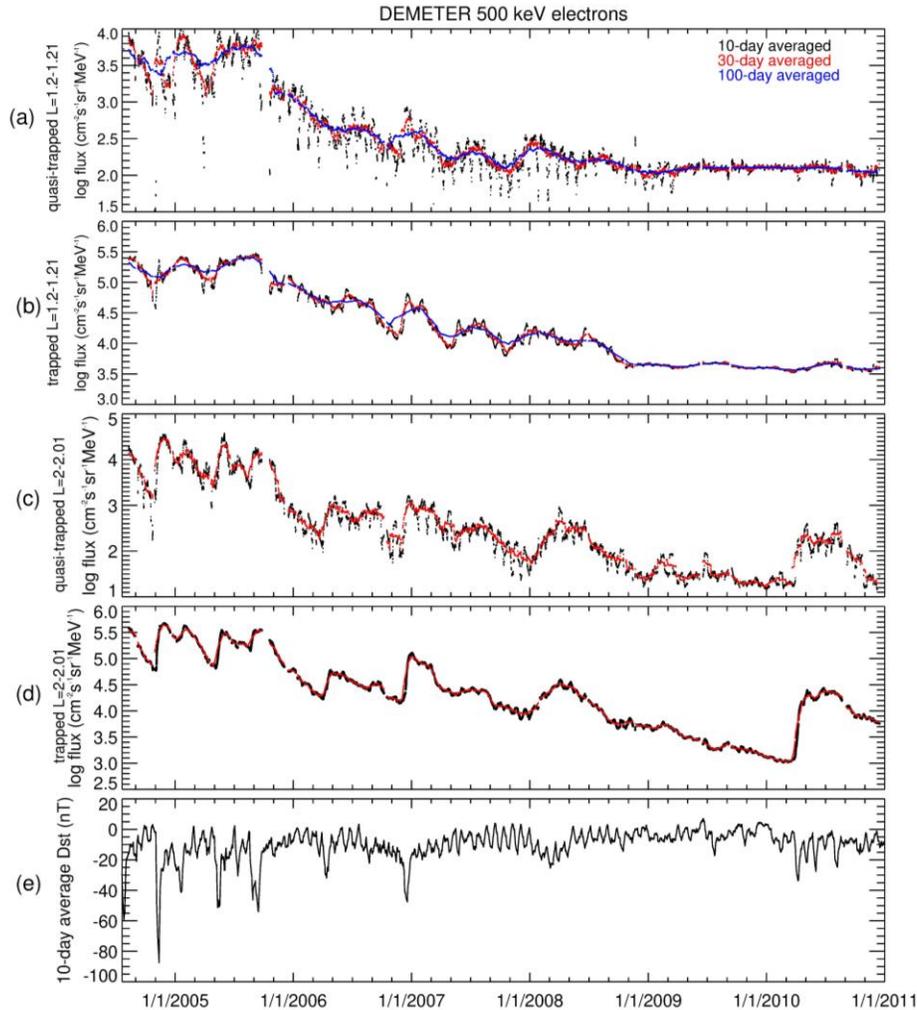

**Figure 7.** Profile of 500 keV quasi-trapped and trapped electrons measured by DEMETER at L=1.2 and L=2 compared with the Dst index. Data are averaged in 10-day (black), 30-day (red) and 100-day (blue) running windows. (a, b) Quasi-trapped and trapped electron fluxes at L=1.20-1.21. (c, d) Similar to (a, b) at L=2.00-2.01. (e) 10-day averaged Dst index.

With the adjustment and the filter applied to SAMPEX data, more uncertainties are involved, and we therefore focus on DEMETER observations during storm times for the more detailed quantitative studies below. First, we identify all geomagnetic storms with a minimum Dst below -30 nT from 2004 to 2010. Each storm period is defined as starting when Dst drops below -5 nT and ending when Dst recovers above -5 nT. Then for each storm period identified, we calculate the average flux of quasi-trapped and trapped electrons at each L. Quasi-trapped electron fluxes are shown to increase with the trapped fluxes in both the inner belt and slot region (L=1.2~2.4) with high correlation coefficients (Figure 8), which indicates that the quasi-trapped electrons originate from the trapped electrons or that they both relate to the same mechanism such as a deep penetration event. Moreover, as L increases from 1.2 to 2.4, the correlation coefficient between the quasi-trapped and trapped electron fluxes increases from 0.74 to 0.97. In the slot region such as L=2 and L=2.4, the pitch angle scattering rate is high due to interactions between electrons and various waves such as plasmaspheric hiss wave, and therefore a majority of the quasi-trapped electrons come directly from the trapped electrons, resulting in the good similarities in their response to storms.

In contrast, wave-particle interactions in the inner belt are significantly less active than in the slot region and pitch angle scattering there is mostly from atmospheric collisions, which are only effective when L is extremely low or the particle has a small pitch angle so that it can reach low altitude. Therefore, pitch angle scattering in the inner belt is generally weaker. In addition, because the loss cone opens up rapidly when approaching extremely low L, the quasi-trapped electrons in the inner belt could also originate from the trapped electrons at higher L which enter the drift loss cone when radially diffused Earthward. However, this process is only effective during extremely large geomagnetic storms and cannot explain the source of the quasi-trapped electrons during small storms or in quiet times. Despite of various competing mechanisms that can potentially explain the enhancement of the quasi-trapped electron flux, it is clear that the correlation between the quasi-trapped and trapped electrons in the inner belt is lower than in the slot region and the connections between these two populations require further investigation. Furthermore, as indicated by the blue line in the top panels of Figure 8, it seems that the trapped electron fluxes are capped with an upper limit level at L=1.2 and L=1.5. Figure 7 shows that these high fluxes are related to the large geomagnetic storms. Since the lifetimes of electrons in the inner belt are long, the enhanced fluxes can last for months, resulting in the observation of similarly high fluxes in many storms. The upper limit in the electron flux at L=1.5, in the heart of the inner belt, is extremely sharp, which is possibly related to instrument problems or proton contamination (Li et al., 2015; Selesnick et al. 2019). More detailed studies are required to fully understand the upper limit flux to the electrons in the inner belt.

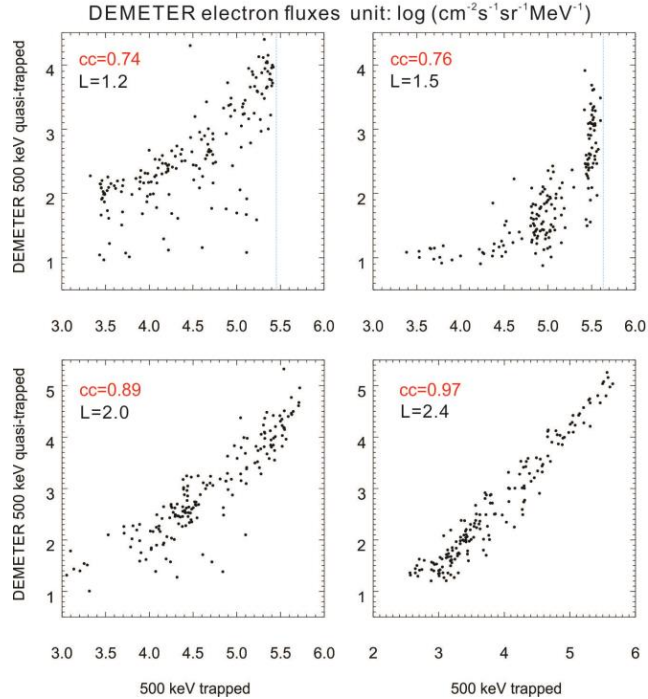

**Figure 8.** Quasi-trapped electron fluxes vs. trapped electron fluxes during storm times at selected L's by DEMETER 500 keV channel. Each black dot represents the average flux of a geomagnetic storm. Correlation coefficients are noted in red. Blue lines in the L=1.2 and L=1.5 panels are illustrations of the upper limit to the trapped fluxes. About 170 events are included in each panel. L range used for the data selection is [L, L+0.01].

**4 Conclusions**

In this study, we investigate the long-term variations and features of inner radiation belt sub-MeV electrons using 7 years of DEMETER and 17 years of SAMPEX data. Our discussions include electrons in the inner belt and slot region from L=1.13 to L=2.4. Measurements from both satellites lead to the following conclusions:

1. Sub-MeV quasi-trapped electrons at the inner edge of the inner belt (L≤1.14) are anti-correlated with sunspot number and positively correlated with cosmic ray intensity indicated by neutron monitor data suggesting their source to be CRAND.

2. The electron flux at the inner edge of the inner belt increased by a factor of two from solar maximum at 2001 to solar minimum at 2009.

3. Both quasi-trapped and trapped electrons at L≥1.2 do not follow the cosmic ray trend, but instead their intensities are enhanced during large geomagnetic storms and decayed during quiet times, indicating sources other than CRAND. During extended quiet times the quasi-trapped electrons decay to a background level that is likely maintained by CRAND especially at L=1.2.

4. Quasi-trapped electron fluxes at L>2 have high correlation coefficients with trapped fluxes, indicating that pitch angle scattering is the dominant source of these quasi-trapped electrons.

**Acknowledgments**

DEMETER data used in this study are publicly available at https://cdpp-archive.cnes.fr. SAMPEX data are available at http://www.srl.caltech.edu/sampex/DataCenter/data.html. Data from the neutron monitor in Mexico City, Mexico are available at http://cr0.izmiran.ru/mxco/main.htm. PAMELA data are from https://tools.ssdc.asi.it/CosmicRays. Dst index is from http://wdc.kugi.kyoto-u.ac.jp/index.html. Sunspot number is from http://www.sidc.be/silso/datafiles. The authors thank Richard Selesnick, Vladimir Mikhailov, Alessandro Bruno and Chen Shi for helpful information and discussion. This work was supported in part by NSF grant AGS 1834971 and NASA grants NNX17AD85G and 80NSSC17K0429 and by NASA/RBSPECT funding through JHU/APL contract 967399 under prime NASA contract NAS5-01072. Z.X. thanks the support from NSFC grant 41904143.